\documentclass[aps,preprint,floatfix]{revtex4}
\usepackage{amsmath}
\usepackage{latexsym}
\usepackage{float}
\usepackage{amssymb}
\usepackage{graphicx}
\usepackage{textcomp}
\usepackage{hyperref}
\def\ba{\begin{eqnarray}}
\def\ea{\end{eqnarray}}
\def\be{\begin{equation}}
\def\ee{\end{equation}}

\def\bm{\begin{math}}
\def\me{\end{math}}

\newcommand{\dummy}

\begin{document}

\title{Pattern, Growth and Aging in a Colony of Clustering Active Swimmers}

\author{Subir K. Das}

\affiliation{\textit{Theoretical Sciences Unit, Jawaharlal Nehru Centre for Advanced Scientific Research, 
            Jakkur P.O., Bangalore 560064, India}
}

\date{\today}

\begin{abstract}
Via molecular dynamics simulations, 
we study the kinetics in a  phase separating active matter model. 
Quantitative results for the isotropic bicontinuous pattern formation,
its growth and aging, studied, respectively, via the two-point equal-time density-density correlation function, 
the average domain length and the two-time density autocorrelation function,
are presented. Both the correlation functions exhibit basic scaling properties,
implying self-similarity in the pattern dynamics, for
which the average domain size exhibits a power-law growth in time. 
The equal-time correlation has a short distance behavior that provides reasonable agreement
of the corresponding structure factor tail with the Porod law.
The autocorrelation decay is a power-law in the average domain size.
Apart from these basic similarities, the quantitative behavior of the
above mentioned observables are found to be vastly different from those of the
corresponding passive limit of the model which also undergoes phase separation. 
The functional forms of these have been quantified. An exceptionally rapid growth in the active system occurs due to
fast coherent motion of the particles, mean-squared-displacements of which
exhibit multiple scaling regimes, including a long time ballistic one.
\end{abstract}
\maketitle

\section{Introduction}

Interest related to phase transitions \cite{onu} in active matters \cite{mar} stems from the fascinating
clustering phenomena \cite{mar} in, e.g.  
a flock of birds, a colony of bacteria, etc. Typically, a steady state in active matter
is the counterpart of the equilibrium in
passive matter.
Phase behaviors \cite{sch,pal,das1,ben}
of various active matter systems have been estimated experimentally
and computationally. Critical exponents have been calculated for simple model systems
\cite{vic,czi,bag,cha}.
Recently, the approach to the steady state, i.e., kinetics of phase transitions,
has been a subject of immense interest \cite{mishra,jul,mishra3,cat,red1,red2,wys,meh,dey,per,mishra2,cre,enys}.

Crucial questions in kinetics \cite{bin,bra,fisd,yeu,maz,cor} relate to
the type of pattern, its growth and aging, as a biological structure builds up from an embryo,
of which, to our knowledge, aging has not been previously studied. 
Interfacial tension and transport mechanism
control these in a passive system.
Motility of the active particles is self-propelling \cite{mar},
examples include microscopic bio-organisms to large animals. Even in a crowded environment this 
can be faster \cite{loi} than diffusion, typical single passive particle dynamics. 
Besides providing a faster collective dynamics, this can
alter the interfacial tension. Thus, the 
kinetics in active matter is rather complex. Specific interest involves the 
scaling properties \cite{bra,fisd}, 
observed in passive matter, and the forms of various correlation and structure functions,
alongside the domain growth law \cite{bra}.
Here we address these issues, for isotropic percolating morphology in space dimension $d=3$, via the
molecular dynamics (MD) simulations \cite{fre,allen} 
of a phase separating model system, having both interparticle interactions 
and self-propelling activity. Existence of a phase transition in the passive limit of the model helps
better quantitative understanding of the effects of activity.

Nature of a pattern can be understood from the two-point equal-time correlation
function \cite{bra}, which, in an isotropic case, is calculated as
$C(r,t)=\langle \psi (\vec{r}, t)\psi(\vec{0}, t)\rangle-\langle\psi(\vec{r}, t)\rangle\langle\psi(\vec{0}, t)\rangle;
~r=|\vec{r}|$.
Here, $\psi$ is a space ($\vec{r}$) and time ($t$) dependent order parameter. 
During the growth, $C(r,t)$ obeys the scaling form \cite{bra}
$C(r,t)\equiv \hat{C}(r/\ell)$,
where $\hat{C}$ is a master function and $\ell$ is the 
average cluster size of a phase separating system. This scaling property reflects
the self-similarity of the pattern during a typical power-law growth \cite{bra},
$\ell \sim t^{\alpha}$.
The exponent $\alpha$ depends upon, among other factors, the transport mechanism \cite{bra}. 
For aging, besides other quantities, one studies the two-time order-parameter
autocorrelation function \cite{fisd}
$C_{\rm{ag}}(t,t_w)=\langle \psi (\vec{r}, t)\psi(\vec{r}, t_w)\rangle-\langle\psi(\vec{r}, t)
\rangle\langle\psi(\vec{r}, t_w)\rangle$,
where $t_w$ ($\le t$) is the age of the system. While
$C_{\rm{ag}}$ decays as a function of $t-t_w$ 
in equilibrium (or in steady state), such time translation
invariance does not hold when $\ell$ has time dependence. During
passive matter transitions, $C_{\rm{ag}}$ has the scaling form \cite{fisd}
$C_{\rm{ag}}(t, t_w) \equiv \hat{C}_{ag}(\ell/\ell_w)$,
$\ell_w$ being the length at $t_w$.

We observe that, in the active case also, both the correlations exhibit the scaling properties,
implying self-similar growth. 
The structure factor, $S(k,t)$, Fourier transform of $C(r,t)$, closely follows the Porod law \cite{porod,oon}.
Other than these,
the structure and dynamics between the active and passive cases are vastly different.
In the active case, even though the growth of $\ell$ is algebraic, value of $\alpha$ is very high,
as in hydrodynamic growth,
due to ballistic particle motion \cite{ben,wan} over
long period, providing an advection-like collective transport.
For $r>>0$, in the positive domain, $C(r,t)$ decays as $\rm{exp}(-r/\ell)$, much slower than the passive case.
The decay of $C_{\rm{ag}}$ is a power-law, differing qualitatively from passive hydrodynamics \cite{ahm,maj2}. 
Despite qualitative similarity of the decay with the Ising model 
kinetics \cite{yeu}, the exponent is different, which, nevertheless, obeys a lower bound \cite{yeu}. 

\section{Model and Method}

In our model, two particles $i$ and $j$, at distance $r=|\vec{r_i}-\vec{r_j}|$ apart,
interact via \cite{das2,mog,pry}
$u(r) = 
U(r)-U(r_c)-(r-r_{c})(dU/dr)_{r=r_{c}},$
where $U(r)=4\epsilon \left[ (\sigma/r )^{12} - (\sigma/r)^{6} \right]$ 
is the standard Lennard-Jones (LJ) potential, 
$\epsilon$, $\sigma$ and $r_c$ ($=2.5\sigma$) being  
the interaction strength,
particle diameter, and a cut-off distance (for faster computation), respectively. 
Phase behavior of this passive model in temperature ($T$) - (number) density ($\rho$)
plane has been studied in $d=3$. The critical temperature ($T_c$) and 
density ($\rho_c$) for the vapor-liquid transition \cite{das2}
are approximately $0.9 \epsilon/k_B$ and $0.3$, $k_B$ being the 
Boltzmann constant.

We introduced the self-propelling activity via the Vicsek interaction \cite{vic,czi}, where
the direction of motion of a particle is influenced by the average
direction ($\vec{D}_N$) of its neighbors, contained inside the radius $r_c$. 
At each instant, every particle gets an acceleration $f_A$ along $\vec{D}_N$. 
This, of course, will 
lift the temperature of the system, even 
in a canonical ensemble and is appropriate only for systems whose
phase behavior is insensitive to temperature \cite{das1,ben}. To avoid this effect, in our temperature controlled
transition, the amplitudes of the velocity vectors, after the kicks,
were normalized to their values prior to the kicks,
so that there is only a directional change.
Other studies report that the Vicsek activity broadens
the coexistence region \cite{das1,ben}. This was observed for this model as well. However,
we will not present results on this aspect.

For this model, we have performed MD simulations \cite{fre,allen} with a Langevin thermostat, by
solving the equation
$m\ddot{\vec{r}}_i=-\vec{\nabla} U_i -\gamma m \dot{\vec{r}}_i + \sqrt{6\gamma k_B T m} \vec R(t)$,
in periodic cubic boxes of linear dimension $L\sigma$,
where $m$ is the mass of a particle, $U_i$ is the potential from interparticle passive interactions,
$\gamma$ is a damping coefficient and $\vec R$ is a noise having $\delta$ correlation in space
and time. A hydrodynamics preserving thermostat
is avoided, since the objective is to see the effects of Vicsek activity.
Presence of fast hydrodynamic mechanism may not allow us to identify the effects of activity.
The Verlet velocity algorithm \cite{fre} was used with time step
$\Delta t=0.01$, in units of $t_0=\sqrt{\sigma^2m/48\epsilon}$. For the
rest of the paper, we set $m$, $\epsilon$, $\sigma$, $k_B$ and $\gamma$ to unity. 
We have prepared the
initial configurations at $T=10$ with $f_A=0$ and kinetics was studied after quenching these to
$T=0.5$ with $\rho=0.3$, for $f_A=0$ (passive) and $1$ (active). 

\begin{figure}[htb]
\centering
\includegraphics*[width=0.6\textwidth]{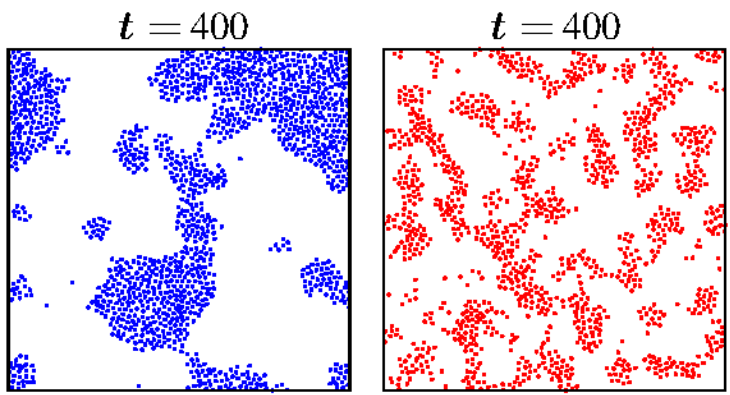}
\caption{
2D cross-sections of the snapshots at $t=400$ 
during the evolutions of the active (left frame) and passive (right frame)
systems. In both the cases, the value of $L$ is
$64$, for which the boxes contain $N=78644$ particles. For kinetics, all the passive model results were
obtained for this particular system size whereas, the rest of the active matter results were
obtained with $L=100$ ($N=300000$).
}\label{fig1}
\end{figure}

\section{Results}

In Figure \ref{fig1} we present 2D cross-sections of snapshots during the
evolutions of the active and the passive systems. Even though a lack of interconnectedness is seen,
in 3D the morphologies are percolating.
Given that the value of $t$ is same in
the two cases, it is clear that the coarsening in the active case
is much faster. Another interesting noticeable difference is in the structure. A possible reason for
this difference is that the Vicsek activity plays the primary role
in the active matter structure formation, unlike the passive case where 
interfacial tension plays a crucial role \cite{bra}.
To have a quantitative knowledge of the difference, we take a look at the $C(r,t)$.

\begin{figure}[htb]
\centering
\includegraphics*[width=0.6\textwidth]{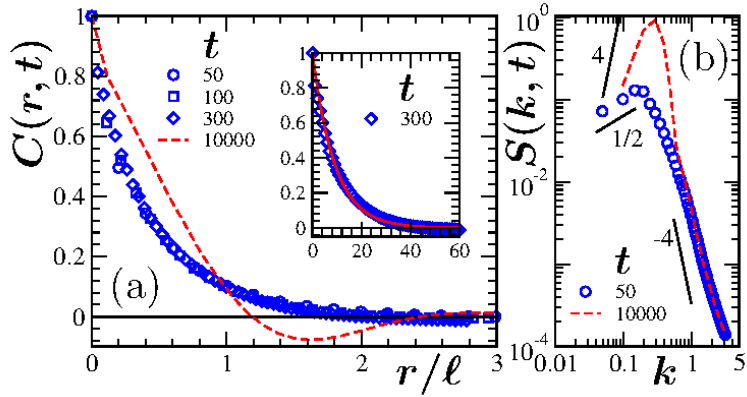}
\caption{(a) Plots of $C(r,t)$ vs $r/\ell$,
using data from different times. Symbols are for active system and the dashed line is for the
passive system. Inset: Plot of $C(r,t)$ vs $r$ for the active system at $t=300$. The solid
line there is a fit to the form $\mbox{exp}(-r/\ell)$ for the data range [0,40]. 
(b) Plots of $S(k,t)$ as a function of $k$ for the active (symbols) and passive (dashed line) systems.
The solid lines are power-laws, the exponents for which are mentioned. The data sets are not normalized.
The ordinates are appropriately scaled to superimpose large $k$ data for the two cases. Here and
for the following results the data were averaged over at least five independent initial configurations.
}\label{fig2}
\end{figure}

In Figure \ref{fig2} (a) we show the scaling plots of $C(r,t)$  
for the active system (symbols). For this calculation,
we have mapped a continuum system onto a simple cubic lattice \cite{das2}. A lattice point was assigned an
order-parameter value $\psi=1$ if the density (calculated using the nearest neighbor region)
at that point is larger than $\rho_c$, otherwise $-1$.
Nice collapse of data from different times confirms the
self-similarity of the growing structure \cite{bra}. Thus, it is meaningful to extract the average
cluster size from the decay of these functions and we obtained it from
$C(\ell,t)=0.1$. 

The dashed line in Fig \ref{fig2} (a) is the $C(r,t)$ for the
passive system. The difference between the two cases is overwhelming, the passive one having 
strong  oscillations \cite{das3}. The absence of this aspect in
the active case has similarity with the $C(r,t)$
for the cluster growth during cooling in an assembly
of inelastically colliding granular particles \cite{das3}.
In the latter system, velocities of the particles become parallel after each collision. The Vicsek
model produces similar parallelization effect and surface tension plays
minimal role in deciding the structure. In that case, long distance behavior of the $C(r,t)$
may have similarity with those for the single phase with large correlation length \cite{fisher}. 
To be on the simpler side and keeping the small $r$ singularity, related to Porod tail 
(a consequence of scattering from the interfaces), in mind, in the inset we use
the form $\mbox{exp}(-r/\ell)$ to fit an unscaled active system $C(r,t)$
in the range $[0,40]$. 
At very large $r$, $C(r,t)$ will have negative values,
before finally decaying to zero, due to the
order parameter conservation constraint \cite{yeu2} $\int C(r,t) d\vec r = S(0,t)=0$.
Thus, while fitting to this form, it is mandatory to exclude data for such extreme limit.
The fit looks certainly good. In fact a good linear behavior can be seen on a log-linear
plot, for $C(r,t)>0.05$, up to $r/\ell$ significantly larger than unity.
This is a much slower decay than even the Ohta-Jasnow-Kawasaki function \cite{bra}
for the nonconserved order-parameter dynamics and is expected to influence the aging property
significantly. 
An Ornstein-Zernike \cite{fisher}
type form provides a better fit, with the power-law exponent being close to zero.

In Figure \ref{fig2} (b) we show the structure factors for the active (symbols) and the passive (dashed line) systems.
In the long wave-vector region there is nice agreement between the two cases and except for
very large $k$ (where the data suffer from noise due to non-zero temperature) 
the tails are reasonably consistent with a decay
(albeit over a narrower range of $k$ in the active case)
$\propto k^{-(d+1)}$, the Porod law \cite{porod} in $d=3$, for a scalar order parameter.
The exponent $\beta=4$ corresponds to the Yeung's law \cite{yeu2}, 
for small $k$ power-law ($k^{\beta}$) enhancement of $S(k,t)$. The
value of $\beta$ appears to be $\simeq 1/2$ (which decreases with time) for the active dynamics,
a consequence of slow decay of the $C(r,t)$, which will be further discussed in the 
context of aging. These results are at variance with the understanding that patterns
are independent of the kinetic mechanism \cite{spuri}. In the passive context,
it was indeed demonstrated \cite{ahm2} that irrespective of a diffusive or a hydrodynamic growth,
$C(r,t)$ remains unchanged. This highlights the complexity of active matter phase separation.

\begin{figure}[htb]
\centering
\includegraphics*[width=0.6\textwidth]{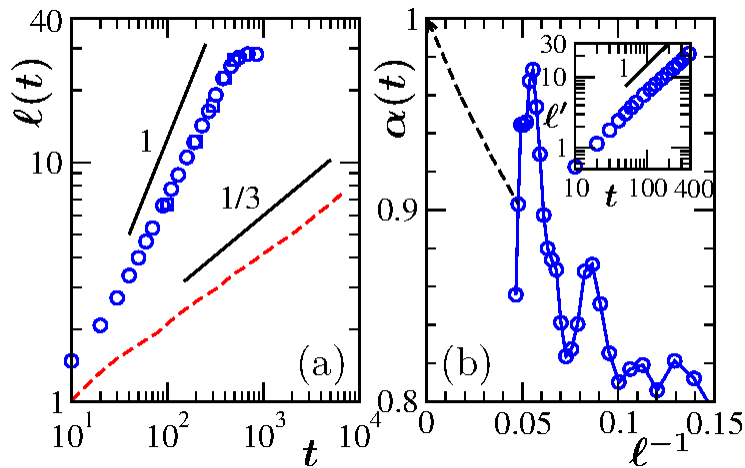}
\caption{(a) Plots of $\ell$ vs $t$, for active (symbol) and 
passive (dashed line) systems. The solid lines correspond to different power-laws,
exponents for which are mentioned. For the active case, the circles correspond
to $C(\ell,t)=0.1$ and the squares (appropriately scaled to superimpose with
the circles) are from exponential fits to the $C(r,t)$.
(b) The time dependent growth exponent for the active system is plotted
vs $1/\ell$. The dashed line is a guide to the eyes. The inset shows
$\ell^{\prime}$ vs $t$ (see text for details).
}\label{fig3}
\end{figure}

In Figure \ref{fig3} (a) we present $\ell$ vs $t$ data, on a log-log scale. For both
type of systems, power-laws are visible. The robustness of the exponential form
for $C(r,t)$ at all times, for the active case, can be appreciated from the fact that the values of $\ell$
extracted from the corresponding fits (squares) are very much parallel to the ones
obtained from $C(\ell,t)=0.1$ (circles). For
the passive case, given that the Langevin thermostat is a stochastic one, diffusive
growth is expected, with $\alpha=1/3$, the Lifshitz-Slyozov exponent \cite{lif}.
The self-propelling activity drastically enhances this value, 
previously observed in
other situations \cite{jul,cre,enys}, after a brief period with $\alpha=1/3$,
to about unity, a behavior expected for a viscous hydrodynamic
growth \cite{sig}. In both the cases, the long time values of the 
exponents appear little weaker than the quoted numbers.
This is due to the presence of a non-zero initial length \cite{maj} $\ell_0$. For more accurate
estimate of $\alpha$ for the active case, 
in Figure \ref{fig3} (b)
we have presented the instantaneous exponent \cite{hus} $\alpha (t)$ ($=d\mbox{ln}\ell/ d\mbox{ln}t$),
as a function of $1/\ell$. Clearly, the exponent is approaching a value $\simeq 1$, asymptotically.
In the inset of this figure, we show the log-log plot for $\ell^{\prime}$ vs $t$, 
where $\ell^{\prime}=\ell-\ell_0$, $\ell_0=0.93$. For a significant range of the
length this data set appear almost perfectly parallel to the $\alpha=1$ line. 
In the passive case, we avoid such exercise, since  
the Ising value of $\alpha$ is expected,
which has been estimated previously via other more sophisticated analyses \cite{maj}.
Our active matter result,
combined with 2D studies \cite{jul,enys}, indicate that the dimensionality dependence may
be weaker than hydrodynamic phase separation in passive matter.

\begin{figure}[htb]
\centering
\includegraphics*[width=0.55\textwidth]{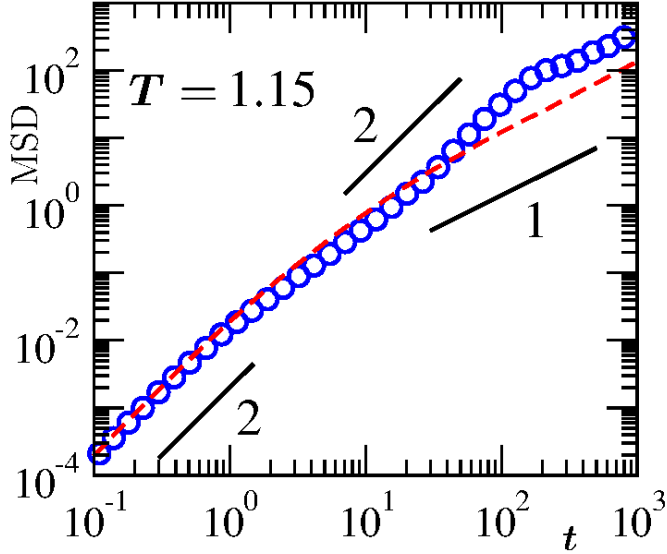}
\caption{Single particle MSDs are plotted vs time. The symbols 
are for the active case, whereas the dashed line is for the passive system.
Unlike the other results, this figure corresponds to $T=1.15$
and $L=24$. The solid lines are power-laws, exponents for which are mentioned.
}\label{fig4}
\end{figure}

Such a fast, hydrodynamic like growth is possible if
the Vicsek activity can produce an advection-like effect in the collective transport.
Given that coherency is inherent in the Vicsek model,
fast particle motion can support that possibility.
In Figure \ref{fig4} we have presented
a comparative picture for the time dependence of the single particle mean-squared-displacements (MSD).
This figure is for $T=1.15$, outside the coexistence
region. The results are very different, despite the fact that 
the passive MSD was obtained via MD runs in a microcanonical ensemble that
perfectly preserves hydrodynamics \cite{fre}. 
While a diffusive behavior for the passive case starts at
$t\simeq 30$, in the active case this is delayed beyond $t=100$. This effect will
be more pronounced inside the miscibility gap. However, there one needs to track particles belonging
to the clustered regions only, making it difficult to obtain data over an elongated period.
Interestingly, compared to the usual scenario, 
there exist multiple scaling regimes \cite{loi} for the active matter MSD. Early time ballistic behavior
crosses over to a super-diffusive regime at an intermediate time scale. Before moving to the
final diffusive regime, a very robust ballistic behavior appears once more. The first two regimes
are connected to the passive interparticle interaction, whereas the long time behavior is due to the 
Vicsek activity.

\begin{figure}[htb]
\centering
\includegraphics*[width=0.6\textwidth]{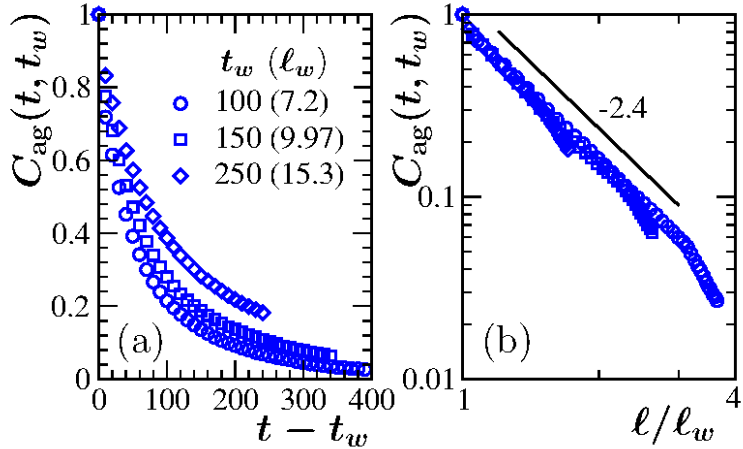}
\caption{(a) Plots of $C_{\rm{ag}}$ vs $t-t_w$, for the active case. Data for three different 
ages are presented. These values, along with the corresponding $\ell_w$, are mentioned.
(b) Log-log plots of $C_{\rm{ag}}$, vs $\ell/\ell_w$, for the same set of $t_w$ values as in (a). The solid line
is a power-law with exponent $\lambda=2.4$.
}\label{fig5}
\end{figure}

Even though advection-like transport property in the active matter
resembles hydrodynamic growth in fluids, these are not exactly the same, since 
the structure in the active case differs from the universal
passive form \cite{spuri,ahm2}. Further differences may be observed in the aging property that we examine next. 
In Figure \ref{fig5} (a), we show the plots
of $C_{\rm{ag}}(t,t_w)$ vs $t-t_w$, for three different values of $t_w$, for the active case. 
As expected, no time translation invariance is obeyed. In Figure \ref{fig5} (b) we plot
$C_{\rm{ag}}(t,t_w)$ vs $\ell/\ell_w$. Nice collapse is seen, similar to the scaling
property in passive transitions. 
Deviations from this, towards the tail of
each data set, are due to the finite-size effects \cite{mid1}. These are expected to appear at smaller
values of $\ell/\ell_w$ for larger $t_w$.
On the log-log scale, 
the robust linear look implies a power-law decay \cite{fisd},
$C_{\rm{ag}}(t,t_w)\sim (\ell/\ell_w)^{-\lambda}$. 
The data in the collapsed region provide 
$\lambda=2.4$. This is remarkably different from the aging in hydrodynamic growth for which
$C(t,t_w)$ decays exponentially fast \cite{ahm,maj2}. Here note 
that, unlike the active case, no 
long range order in the velocity field was observed \cite{ahm2} even in (passive) hydrodynamic phase separation.

Prediction exists \cite{fisd,yeu} for
a lower bound on $\lambda$, viz.,  $\lambda\ge (d+\beta)/2$. Since
$\beta$ is significantly less than even unity, due to very slow decay of $C(r,t)$, the observed
value, viz., $2.4$, is consistent with this bound. 
This number, however,
is even much smaller than the corresponding passive case \cite{mid2} with diffusive kinetics, 
by accepting that the passive result
will be same as the conserved Ising model. The difference in the values of $\beta$ is
partly responsible for this. 
The rest can be attributed to the fact that in the
passive case domains can have fair degree of random motion, enhancing
loss in memory. This effect should be more prominent in fluids, that leads to the exponential fall in
$C_{\rm{ag}}(t,t_w)$.
On the other hand,
even though fast hydrodynamic-like growth occurs due to coherent dynamics of the ballistically moving
particles in active matter, such motion makes the movement of the domains more deterministic. In fact, it is reported
\cite{enys} that the domains can slide along the interface, supporting our
argument. 
Nevertheless,
there exists scope for better understanding of this issue.

\section{Summary}

In conclusion, we have obtained a quantitative picture of the influence of self-propulsion on
various aspects of kinetics of phase separation,
viz., structure \cite{bra}, growth \cite{bra} and aging \cite{fisd}, in a model of active matter,
 for bicontinuous morphology.
The fundamental properties
of phase ordering dynamics are obeyed, namely the scalings in the two-point equal-time
and two-time autocorrelation functions are observed, implying a self-similar growth \cite{bra}.
However, there exists significant quantitative difference between the active and passive matter kinetics.
A hydrodynamic like behavior \cite{sig} in the active matter domain growth
is observed due to the fast coherent ballistic motion of the particles, leading to an
effective advection in the collective dynamics. This, interestingly, makes the 
structure deviate from corresponding passive universal behavior.
The correlation function
shows nearly an exponential decay, much weaker than the passive case, except for small $r$.
Consequently, the decay of the autocorrelation is also much slower than similar passive systems, though
obeys a lower bound \cite{yeu2}. 

{\bf Acknowledgment:} The author acknowledges comments from K. Binder and
financial supports from Department of Science and Technology,
Government of India; Marie-Curie Actions Plan of European Commission (FP7-PEOPLE-2013-IRSES
Grant No. $612707$, DIONICOS) and International Centre for Theoretical Physics, Italy.

\end{document}